# Multiple Service providers sharing Spectrum using Cognitive Radio in Wireless Communication Networks

R. Kaniezhil and Dr. C. Chandrasekar

**Abstract**— The current utilization of the spectrum is quite inefficient; consequently, if properly used, there is no shortage of the spectrum that is at present available. Therefore, it is anticipated that more flexible use of spectrum and spectrum sharing between radio systems will be key enablers to facilitate the successful implementation of future systems. Cognitive radio, however, is known as the most intelligent and promising technique in solving the problem of spectrum sharing. In this paper, we consider the technique of spectrum sharing among users of service providers to share the licensed spectrum of licensed service providers. It is shown that the proposed technique reduces the call blocking rate and improves the spectrum utilization.

**Index Terms**— , call blocking, CR, service provider,spectrum sharing, spectrum utilization,system efficiency.

———————————— ◆ ————————————

## 1 INTRODUCTION

Nowadays, there are a lot of wireless applications sharing the same medium. This overload leads to a lack of spectrum in given frequency bands. At this time, the allocation of the spectrum is mostly static and based on licensing, as in the case of the FM-band. The purpose of this work is to present another approach of sharing the same medium by means of a dynamic allocation of the spectrum instead of the static allocation which is believed to cause spectrum inefficiency and scarcity, since the wireless users' demands change both temporally and spatially.

Dynamic spectrum sharing is a promising approach for reusing the underutilized spectrum, in which the spectrum is shared among primary and secondary (unlicensed) users (SUs) to improve spectrum flexibility and, therefore, efficiency. The cognitive radio (CR) provides a solution to this dynamic spectrum access (DSA). CR technology is a key enabler for both real time spectrum markets and dynamic sharing of licensed spectrum with unlicensed devices. The CR device is able to perform spectrum acquisition, either through purchasing (in cleared spectrum) or sensing (in vacant channels of geographical interleaved spectrum), over a range of frequency bands and operate in this spectrum at times and locations where it is able to transmit in a non-interfering basis.

The overall spectral efficiency of a system can be improved with good coexistence properties, good spectrum sharing capabilities, as well as with flexibility in the spectrum use. Capabilities to share spectrum with other systems will significantly increase the efficiency as well as acceptability of the system. The overall spectral efficiency of the work can be also increased with a flexible use of spectrum that adapts to the spatial and temporal variations in the traffic and environment characteristics. The flexibility and scalability of the system is important also in order to simplify the network deployment under spectrum arrangements that may vary from region to region. Built-in capabilities for flexibility and sharing may significantly ease the task of spectrum identification for the system.

In this paper, we propose a scheme about the efficient spectrum utilization (ie spectrum sharing) among service providers via static CR nodes where the under utilized spectrum of a particular service provider can be shared by the overloaded service providers with a coordination among them. This also proposes deploying a network of fixed cognitive radio (CR) nodes to maintain spectrum sharing across multiple service providers operating in the same geographical area. These CR nodes estimate the spectrum utilisation in a given area, and cooperate to provide the spectrum usage information for the overloaded infrastructures of service providers. The algorithm SBAC is set to the beyond 3rd generation (B3G) wireless communication systems.

I have already proposed the Minimizing the Interference and Performance of QoS based on spectrum sharing using CR nodes[1] and the overall spectral efficiency is carried out in the present work.

The paper is organized as follows; Section 2 and 3 defines cognitive radio and proposes a system model approach for its implementation. In section 4, Performance analysis has been investigated to improve the system efficiency. Section 5 presents the simulation results and implementation issues. Finally, conclusions are presented in Section 6.

## 2 COGNITIVE RADIO

The key enabling technology of dynamic spectrum access techniques is cognitive radio (CR) technology, which provides the capability to share the wireless channel with licensed users in an opportunistic manner. The term, cognitive radio, can formally be defined as follows :

A "Cognitive Radio" is a radio that can change its transmitter parameters based on interaction with the environment in which it operates. From this definition, two main characteristics of the cognitive radio can be defined as follows:

- **Cognitive capability**: It refers to the ability of the radio technology to capture or sense the information from its radio environment. Through this capability, the portions of the spectrum that are unused at a specific time or loca-

—————————————————

- R. Kaniezhil is currently pursuing Ph.D in Computer Science, Periyar University, Salem, India, PH-9994451525. E-mail:kaniezhil@yahoo.co.in.

tion can be identified. Consequently, the best spectrum and appropriate operating parameters can be selected.

**Reconfigurability**: The cognitive capability provides spectrum awareness whereas reconfigurability enables the radio to be dynamically programmed according to the radio environment..

CR networks are envisioned to provide high bandwidth to mobile users via heterogeneous wireless architectures and dynamic spectrum access techniques. This goal can be realized only through dynamic and efficient spectrum management techniques.

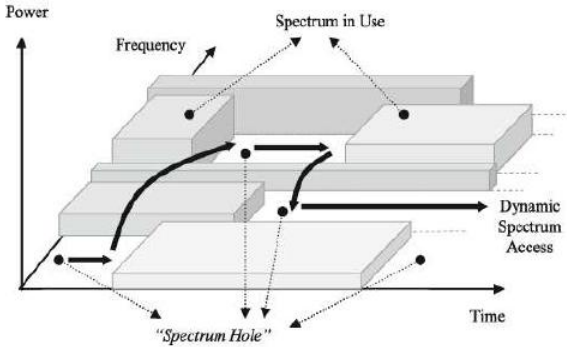

Fig. 1. Spectrum Hole Concept

The ultimate objective of the cognitive radio is to obtain the best available spectrum through cognitive capability and re-configurability as described before. Since most of the spectrum is already assigned, the most important challenge is to share the licensed spectrum without interfering with the transmission of other licensed users as illustrated in Fig 1.

The cognitive radio enables the usage of temporally unused spectrum, which is referred to as spectrum hole or white space. If this band is further utilized by a licensed user, the cognitive radio moves to another spectrum hole or stays in the same band, altering its transmission power level or modulation scheme to avoid interference as shown in Fig 1.

### 2.1 Proposed CR Nodes Sensing

With the capability of sensing the environment and finding the available spectrum dynamically, CR technique can help to implement spectrum sharing and improve the spectrum utilization efficiency. In the proposed work, CR engines such as learn and decision does the work of Spectrum Sensing. Here, CR nodes helps to provide the list of available channels and it checks the availability of the channels. Mobile nodes get the information of availability of channels and sends it to BS through the CR nodes as shown in Fig 2.

When CR nodes monitor channels in a given area, economically, the number of deployed CR nodes should be minimal, and, meanwhile, these CR nodes can fully cover the given area and properly estimate the channel utilisation.

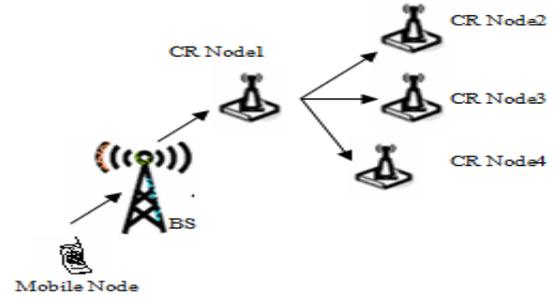

Fig.2 CR Nodes sensing the available channels

## 3 Proposed Spectrum Sharing Techniques

### 3.1 Process of Specrtum Sharing

During the peak hours, the communication of the users will be blocked due the number of active users is greater than the maximal number of users ie the infrastructure is found to be overloaded(because of the channel scarcity). At the same time, infrastructure of the other service providers might be in the under-loaded status. Hence, the available channels of the under-loaded service providers can be utilized by the overloaded service providers.

This may vary according to the services offered by the customers example Wireless internet for laptops, mobile communications and so on. So this may lead to the difference in the traffic of the active users across service providers. Here, both of these two service providers operate cell-based wireless networks. Therefore implementing spectrum sharing among service providers would highly improve the spectrum efficiency and it also reduces the call blocking rate and Co-Channel Interference.

In order to reduce the Co-Channel Interference and to remove the need of equipping CRs in each infrastructure of service providers, we propose a method that implements a spectrum sharing among service providers via CR nodes in a long term spectrum assignment scheme. These CR nodes are distributed regularly within an area of interest. Each CR node senses the surrounding environment and monitors the channel usage within its sensing range of different service providers.

To avoid co-channel Interference, CR nodes provides the list of the channel availability of each cell for the overloaded infrastructure of service provider. CR nodes are connected to each other via wire or they communicate wirelessly to form a network. This network is called as spectrum management network and it coexists with the wireless networks operated by different service providers.

In this technique, only a limited number of CR nodes are required, and neither users nor service providers need to sense the environment for available spectrum. In the proposed technique, each user subscribes to a specific service provider that is assigned fixed frequency bands. When one or more infrastructures (e.g. base stations) of a service provider are overloaded, they use extra available channels (for communication) which are licensed to other service providers. The overloaded infrastructures obtain the channel availability information from surrounding CR nodes. CR nodes are deployed to estimate the channel utilisation and provide the channel usage information for infrastructures upon their requests. The infra-structures process the information received from CR nodes to select the optimum channels based on the channel associated metrics such as interference level, cost and the probability of channel being available for a certain time duration.

### 3.2 Operations of CR Nodes

Suppose, if the infrastructure of one service provider is found to be overloaded, it sends the request to the adjacent CR nodes regarding the channel usage availability in its coverage area. It gives the list of available channel from the relevant adjacent CR nodes and it also provides the relevant information associated with each channel such as the average signal to noise–plus interference.

The overloaded infrastructure now selects the proper channel to use. Thus, users can communicate with the overloaded infrastructure over the new channels after they are informed with the channel availability. This requires both infrastructure and users to be equipped with radios capable of operating over different frequency bands. If the infrastructure is not found to be overloaded or the traffic is free, then the channels should be released ie users would stop using these channels.

The main operation of CR nodes is to periodically sense the environment and estimate the channel utilization within their sensing range. Once it receives the channel availability information from infrastructures, CR nodes would response to these infrastructures with a set of available channels and relevant information. The main challenges that we facing during the spectrum sharing are:

(i) How to decide the availability of a channel – for a CR node ?

(ii) How to select the optimum channels for usage based on the information provided by CR nodes – for the infrastructure?

The solutions are given by the determination of channel availability in CR nodes and in an infrastructure ie CR nodes senses the unused channels and an overloaded infrastructure would receive the channel usage information from its adjacent CR node after request. This can be proceeded by the SBAC Algorithm(Selection of Best Available Channel).

Thus, implementing those CR nodes and deploying them as a network should be done with the negotiation between service providers based on the cost and the network management policy. This successful deployment provides the risk and the cost of operating the CR network. It founds to be a challenging task of deploying the CR network with the coexistence with the current wireless networks in a geographical situation. It also increases the cost and complexity of wireless network management. If CR nodes communicate with each other wirelessly, then it requires extra wireless resources and it also increases the overhead of wireless networks.

---

### Algorithm : SBAC

$BS \leftarrow mn$ (mobile node send request to Base Station)

$CR \leftarrow BS$

$nCR \leftarrow CR$

$current\_channel\_available\_list \leftarrow nCR$

$prob \leftarrow current\_channel\_available / total\_channel$

$\forall ch\_frq$

 if $frmax < ch\_frq$

  $frmax \leftarrow frq$

 end

 if $frmin > ch\_frq$

  $frmin \leftarrow ch\_frq$

 end

end

$inter \leftarrow | frmax - frmin |$

$cost \leftarrow t * 60 * c$

$ch\_u = (10 * \beta_1 * prob) + \beta_2 * \log\frac{1}{inter} + \beta_3 * \frac{1}{cost}$

$cu\_list[cu\_count++] = | ch\_u |$

if $(cu\_count != 0)$

 $\forall cu\_list$

  if $maxi < cu\_list$

   $maxi = ch\_list$

  end

 end

$channel\_maxi \leftarrow maxi$

end

---

### 3.3 Simulation Results of Spectrum Sharing

If the infrastructure of one service provider is found to be overloaded, it sends the requests to the adjacent CR Nodes regarding the availability of Channel. Channel availability can be determined by sending service request to the BS. BS receives service request from the mobile nodes and it will send the channel request to the CR node.Fig.3 shows the BS before sending request to the CR Nodes.

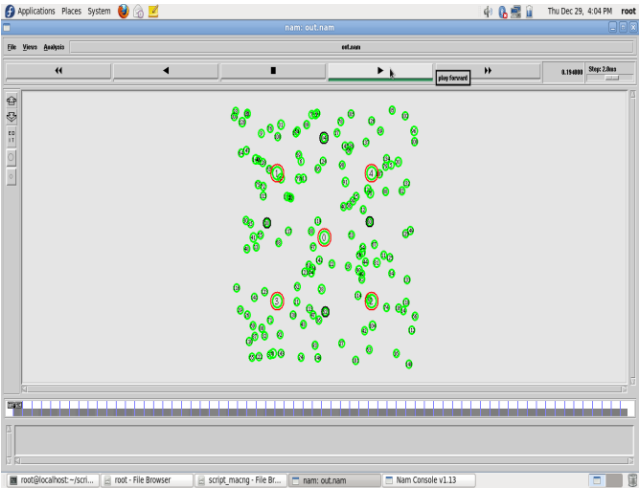

Fig. 3 Before sending request to the CR Nodes

CR node receives the channel request and sends the broadcast message to the adjacent CR node. A neighbor CR node receives the broadcast message and also sends the available channel list to the BS. CR node and its neighbors, updates the channel availability list and sends response to the BS. Fig.4 shows the sending and receiving the request and response from CR nodes.

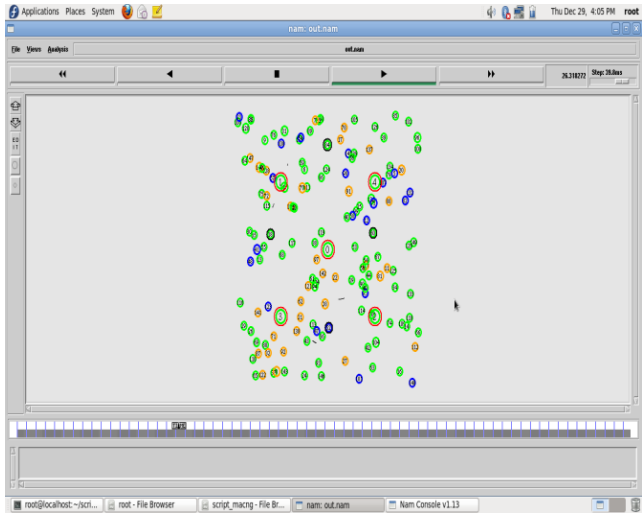

Fig.4 Sending request to the CR nodes and response from the CR nodes.

If BS receives the response from the CR node, it selects the available channel and sends service reply with the allocated channel to the mobile nodes. This shows the maximize utilization of a channel and it offers several services such as internet service, call service, multimedia service and so on to the mobile nodes. Fig.5 shows the maximum utilization of channel by sending the response to the BS regarding the channel availability.

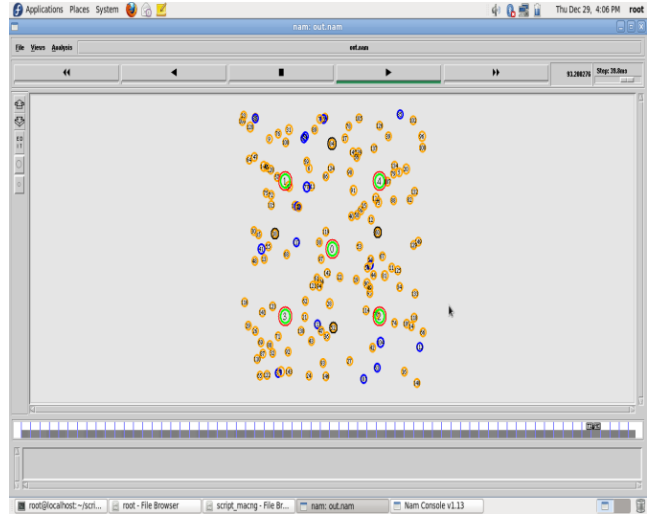

Fig.5 Response from CR nodes

Our scheme, identifies 'available' channel list for each CR node. Such a list shows which channel is available to use depending on the distance among the CR node and High frequency band. Within a given neighborhood, multiple CR nodes may contend for access to one or more of the available channels.

## 4 PERFORMANCE METRICS

In this section, we discuss about the performance metrics study the impact of spectrum sharing on the service providers include call blocking rate, system efficiency, Revenue efficiency, etc.

### 4.1 Call Blocking Rate

The call blocking rate $R_{BL}$ is defined as the ratio of total blocked calls over total calls processed by all service providers and corresponds to:

$$R_{BL} = \lim \frac{n_{BL}^{(total)}(t)}{n_{processed}^{(total)}(t)}$$

where total blocked calls at time t by all service providers is given by

$$n_{BL}^{(total)}(t) = \sum_{i=1}^{n_{sp}} n_{BL}^{(i)}(t)$$

and the total calls processed is:

$$n_{processed}^{(total)}(t) = \sum_{i=1}^{n_{sp}} n_{processed}^{(i)}(t)$$

where $n_{sp}$ is the number of service providers. Here, the call would be blocked, if all the service providers are over-loaded.

### 4.2 System Efficiency

The spectrum efficiency is defined as the number of channels used at time $t$ for service provider, and the number of total channels owned by the service provider. Higher spectrum efficiency is anticipated compared to service provider, because the call blocking rate of a user-central system is lower; thus, more calls can contribute to the spectrum utilization.

### 4.3 System Utilization Efficiency

The Spectrum Efficiency $\eta_s^{n_{(sp)}}$ is defined as the ratio of average busy channels over total channels owned by service providers. It corresponds to

$$\eta_s^{n(sp)} = \lim \frac{1}{t} \int_0^t \frac{n_{busy}^{n(sp)}(t)}{N_{ch-total}^{n(sp)}(t)} dt$$

where $n_{busy}^{n_{(sp)}}(t)$ is the number of channels used at time t for service provider $n_{(sp)}$ and $N_{ch\text{-}total}^{n_{(sp)}}(t)$ is the total number of total channels owned by service provider $n_{(sp)}$. Higher Spectrum efficiency is estimated because the call blocking rate is lower; thus more calls can contribute to the spectrum utilization.

### 4.4 Revenue(Cost) Efficiency

Within the observation time, cost is determined by the number of processed calls and the length of call holding time. We define the metric $c_e^{(i)}$ to reflect the cost efficiency. $c_e^{(i)}$ is the ratio of the cost earned within the observation time t over total input traffic intensity for service provider sp, is defined as

$$c_e^{(i)} = c^{(i)}/E^{(i)} = \alpha^{(i)}.E_p.t^{(i)}/E^{(i)} = \alpha^{(i)}.t^{(i)}.\eta_s^{(sp)}$$

where $\alpha^{(i)}$ is the unit price (\$/second/channel) for service provider sp and $c^{(i)}$ is the average income within the observation time.

## 5 SIMULATION RESULTS

In this section, we present simulation results on the performance of our proposed sensing framework. Channel assignment mechanisms in the traditional multi-channel wireless networks typically select the 'Best' channel for a given transmission. In the proposed work, we are choosing the available channel with the high probability and high-frequency band. To generate utility performance measures, we assume:

1) Maximal five service providers share their spectrum, and 100 nodes are chosen. Maximum limit of user per channel is 10.

2) Call arrival of each service provider is the heterogeneous process.

3) Traffic rates are correlated jointly-Gaussian random variables.

4) The infrastructures f different service providers are located at the same position and the cell radii is also the same.

5) The CR nodes are present at the vertices of the cells of the service providers.

6) Each CR node has the ability of sensing its range within the coverage limits.

7) CR nodes have the capability of detecting all the available channels that are licensed to the other service providers.

8) Channel parameters such as interference level, the probability of being available for a given time period and cost are the same for all available channels.

We conduct simulations to verify the potential of the call arrival rate for different service providers in terms of utility performance measures.

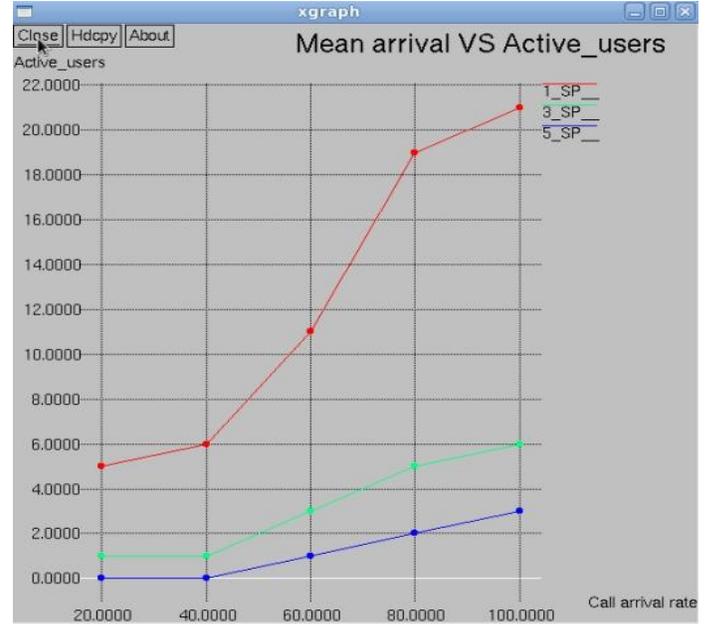

Fig.6  Mean arrival vs Active users

Fig.6 shows that, the active users are reduced when there is a higher traffic in call blocking of different service providers are highly correlated. If the call blocking is reduced among the correlated service providers, then there would be increase in the active users.

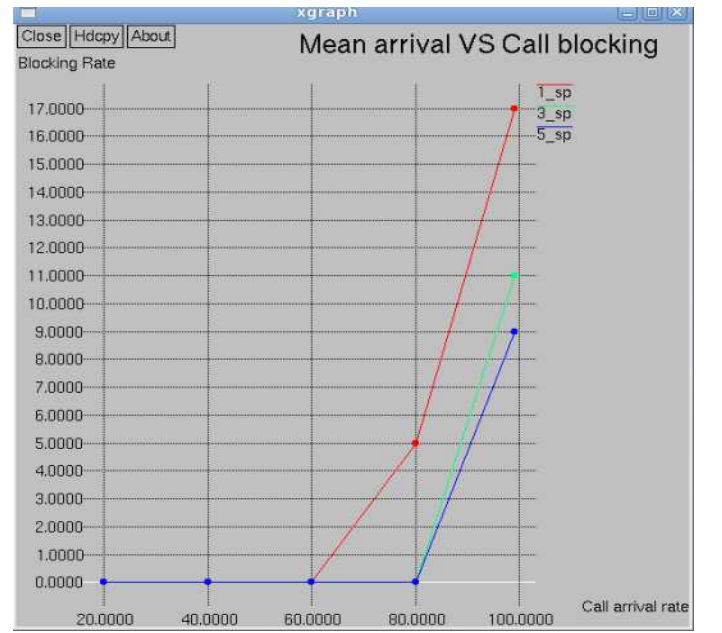

Fig.7  Mean arrival vs Call blocking

Fig.7 shows that at higher traffic rates, the call blocking rate is higher when the traffic rates of different service providers are highly correlated.

Fig.8 shows that as the call blocking is reduced then the traffic load for five service providers is minimum.

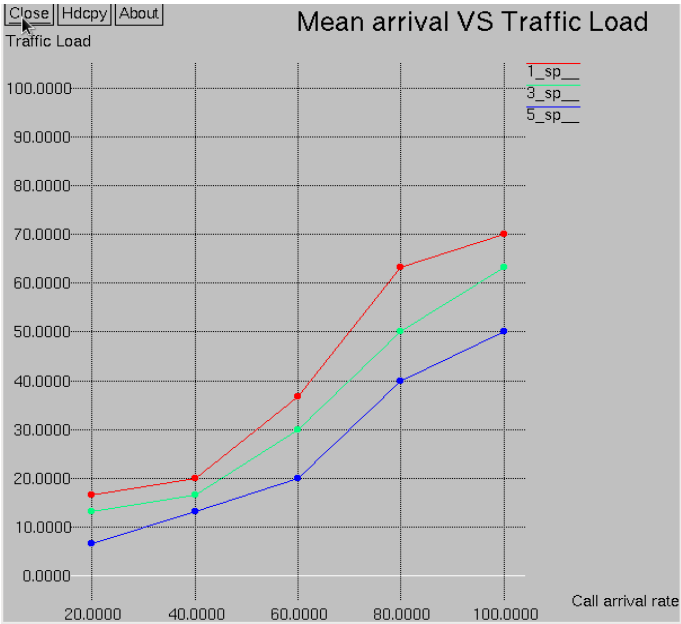

Fig.8  Mean arrival vs Traffic Load

As the mean call arrival increases the channel utilization also increases as in the Fig.9. Higher Spectrum efficiency is estimated because the call blocking rate is lower; thus, more calls can contribute to the spectrum utilization.

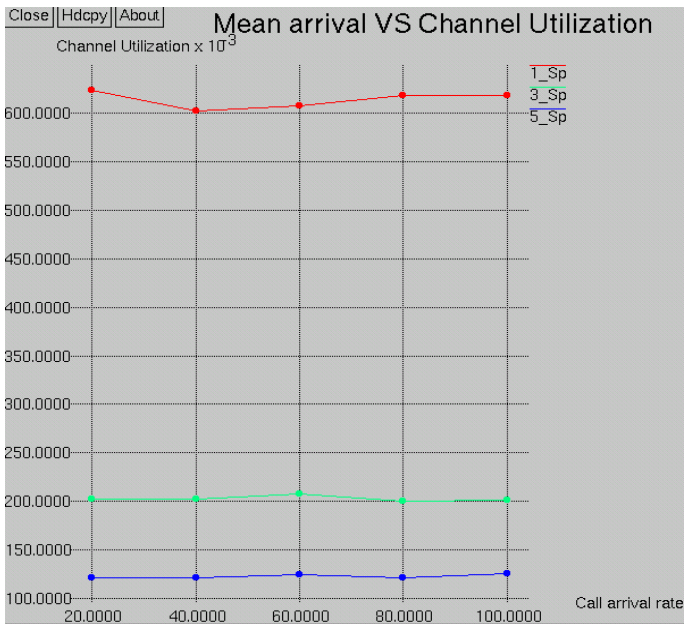

Fig.9 Mean arrival vs Spectrum Efficiency

Fig.10 shows that, at high-traffic rates, the system efficiency is lower when the traffic rates of different service providers are highly correlated. When the correlation is lower, based on Fig.10, as the dropped calls decrease, thus, the total processed calls increase. The system efficiency decreases when the traffic rate is beyond the system capacity.

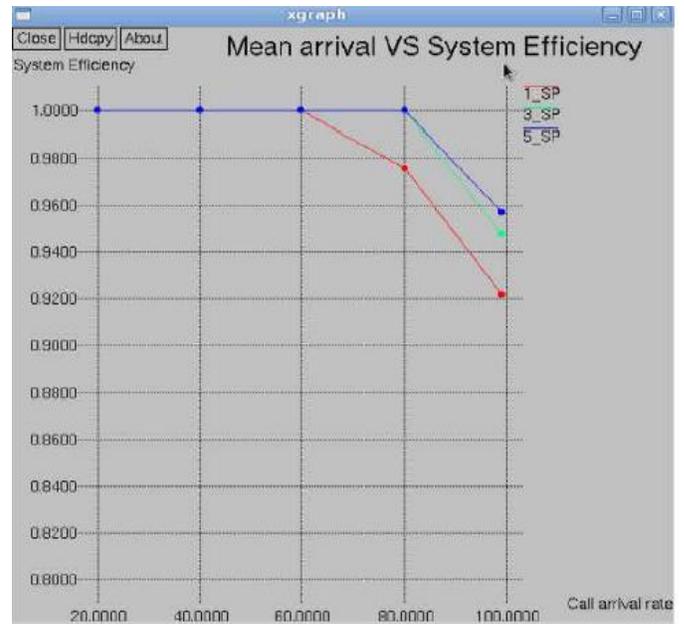

Fig.10  Mean arrival vs System Efficiency

# 6  CONCLUSION

The Spectrum assigned to different service providers is not properly utilized with the same frequency. So some service providers may try to use the allocated spectrum fully, and even they need more spectrums and, which may not be used by service providers fully. This paper provides an offer (way) to utilize and also to share the licensed spectrum among the service providers if they are under utilized. Here, we discussed about the operations of CR nodes and infrastructures of service providers for spectrum sharing. From the proposed algorithm, we can sense the range of CR node and we decide and select the optimal channel for spectrum sharing. In addition to this work, we also define some performance metrics for spectrum sharing among service provides in order to efficient spectrum utilization. This technique removes the need of sensing spectrum in each user. Thus, it reduces the cost, complexity and battery power consumption of user devices.